\documentclass[a4paper,twocolumn,nobibnotes,11pt]{quantumarticle}

\pdfoutput=1
\usepackage[utf8]{inputenc}
\usepackage[english]{babel}
\usepackage[T1]{fontenc}
\usepackage{amsmath, amssymb}
\usepackage{hyperref}

\usepackage{physics}
\usepackage{bm}
\usepackage{graphicx}
\usepackage{verbatim}
\usepackage{subfiles}

\begin{document}

\title{Hardware-efficient formulation of molecular cavity-QED Hamiltonians}

\author{Francesco Troisi}
\email{francesco.troisi@mpsd.mpg.de}
\affiliation{Max Planck Institute for the Structure and Dynamics of Matter and Center for Free-Electron Laser Science, Luruper Chaussee 149, 22761, Hamburg, Germany}

\author{Simone Latini}
\email{simola@dtu.dk}
\affiliation{Department of Physics, Technical University of Denmark, 2800 Kgs. Lyngby, Denmark}

\author{Heiko Appel}
\affiliation{Max Planck Institute for the Structure and Dynamics of Matter and Center for Free-Electron Laser Science, Luruper Chaussee 149, 22761, Hamburg, Germany}

\author{Martin Lüders}
\affiliation{Max Planck Institute for the Structure and Dynamics of Matter and Center for Free-Electron Laser Science, Luruper Chaussee 149, 22761, Hamburg, Germany}

\author{Angel Rubio}
\email{angel.rubio@mpsd.mpg.de}
\affiliation{Max Planck Institute for the Structure and Dynamics of Matter and Center for Free-Electron Laser Science, Luruper Chaussee 149, 22761, Hamburg, Germany}
\affiliation{Initiative for Computational Catalysis (ICC), The Flatiron Institute, 162 Fifth Avenue, New York, NY 10010, United States}

\author{Ivano Tavernelli}
\email{ita@zurich.ibm.com}
\affiliation{IBM Quantum, IBM Research Europe - Zurich, 8803 Rüschlikon, Switzerland}

\begin{abstract}
Light–matter coupled Hamiltonians are central to cavity materials engineering and polaritonic chemistry, but are challenging to simulate with classical hardware due to the scaling of the Hilbert space with the number of quantum photon modes and matter complexity.
Leveraging the fact that quantum computers naturally represent photonic modes efficiently, we present a novel approach to simulate quantum-electrodynamical (QED) systems on near-term quantum hardware. 
After developing the bosonic and mixed operators in the \textit{Qiskit Nature} framework, we employ them to simulate a first-order Trotterized Hamiltonian for a spontaneous-emission problem of a two-level system in an optical cavity.
We find that using a standing-waves photonic basis approach leads to fidelity issues due to hardware connectivity constraints and two-qubits gates errors.
Hence, we propose using a localized photonic basis approach that enforces nearest-neighbor couplings, thanks to which we can map the Hamiltonian as a 1D qubit chain.
We significantly reduce the noise and, by applying the zero-noise extrapolation error mitigation technique, we recover the accurate quantum dynamics.
Finally, we also show that this approach is resilient when relaxing the 1D qubit chain approximation.
\end{abstract}

\maketitle

\section{Introduction}\label{sec:intro_sec}
Over the past decade quantum computers have evolved from proof-of-concept devices with only a few noisy qubits to programmable machines comprising hundreds of qubits.
Current state-of-the-art processors, mostly based on superconducting circuits~\cite{Kim2023}, but also ion-trap~\cite{Wright2019} and neutral-atom~\cite{Graham2022} arrays, have demonstrated problem-specific quantum advantage on carefully chosen problems, such as the Ising model~\cite{Kim2023} or sampling the output of a random quantum circuit~\cite{Arute2019}.
Focusing only on superconducting hardware, technological advancements such as high-fidelity gates, native mid-circuit measurement and advanced error-mitigation protocols~\cite{Temme2017, Endo2021}, allowed the community to shift its focus from hardware demonstrations to physically relevant applications such as protein chain optimization~\cite{Agathangelou2025} or hadron scattering observation~\cite{Schuhmacher2025}.
However, since even state-of-the-art superconducting layouts are still quasi-planar, most previous works dealt with Hamiltonians that map naturally onto a linear chain of qubits, thereby avoiding or minimizing SWAP operations. Examples include Fermi-Hubbard models~\cite{Kokail2019, Khodaeva2024}, Ising~\cite{Kim2023} and molecular Hamiltonians~\cite{Rossmannek2023}.
Systems that mix particles of different nature, such as fermions and bosons, often feature star-like or all-to-all couplings which exceed the capabilities of current devices in terms of connectivity and efficiency.

Materials cavity quantum electrodynamics (QED) provides an example for such systems. Optical cavities confine the electromagnetic field in a small region, allowing to reach a strong light-matter coupling regime with embedded condensed matter systems~\cite{Flick2017, Ruggenthaler2023, Lu2025}.
Tuning the cavity frequencies can modify molecular polarizabilities~\cite{Sidler2024, Horak2025}, modify inter-molecular properties~\cite{SpinGlass}, open gaps in Dirac materials~\cite{CavityGraphene} or reshape excitonic spectra~\cite{cavity_control, Troisi2025}. This emerging field is known as cavity materials engineering~\cite{Lu2025, Hubener2024}.
Classical simulations of QED systems are costly due to the memory required to represent the full Hilbert space as well as to the time required to diagonalize the Hamiltonian, especially when dealing with many photonic modes.
In fact, the photonic Hilbert space scales exponentially with the number of modes and the number of Fock states used to represent them.
Hence, one typically describes a simple matter system coupled to many photonic modes~\cite{Flick2017} or relies on a single-effective (or few-effective) mode description of the electromagnetic field~\cite{Troisi2025, qed_hamiltonian}.
On a quantum processor, by contrast, the same exponential structure can be encoded with a linearly scaling number of qubits. Combined with the short execution time of a quantum circuit, this promises to be a breakthrough in the simulation of QED systems.

As a first step towards a full quantum simulation where both light and matter retain their full complexity, in this work we study the quantum dynamics of a simple fermionic system (two-level system) coupled to a bath of photonic modes. 
First, we present the technical implementation of the Bosonic Operator and Mapper, required to represent the photonic modes on the quantum computer, and of the Mixed Operator and Mapper, 
which enable the representation of fermion–boson interactions, within the open-source Python package \textit{Qiskit Nature}~\cite{QiskitNature}. 
\textit{Qiskit Nature} is part of \textit{Qiskit}, IBM's open source quantum Software Development Kit (SDK)~\cite{qiskit2024}.
Subsequently, we study the time evolution of the aforementioned system, observing the Rabi oscillations. After representing the Hamiltonian on the quantum hardware and defining the initial state such that the matter systems starts in the excited state and all photonic modes in the vacuum, we simulate the quantum dynamics with the Trotter time evolution algorithm. 
In our approach, the matter system is described directly in terms of fermionic operators rather than spin particles. This represents a key novelty compared to previous works on the topic~\cite{Miessen2021} and enables the extension of the method to more complex matter systems.
Moreover, in this work we employ a Trotter approximation of the time evolution operator instead of a variational quantum approach, and we focus on demonstrating the system’s scaling with an increasing number of modes rather than limiting the analysis to only a few modes.
Furthermore, we also acknowledge a previous work on classical hardware~\cite{VacuumFluctuations} that studies the dynamics of a two-level fermionic system coupled to a bath of photonic modes, in particular comparing an exact quantum simulation to a multi-trajectory Ehrenfest dynamics approach.
We present two formulations of the QED Hamiltonian, adapting it to the strong noise requirements of near-term quantum computers, and we show that, even with the presence of SWAP operations, we can extrapolate the noiseless dynamics for one of them, which paves the way for quantum simulations of cavity quantum electrodynamics.

The paper is organized as follows: Section~\ref{sec:methods} presents the theoretical model and the implementation details, Section~\ref{sec:results} discusses the two formulations of the QED Hamiltonian as well as the noisy simulations. Finally Section~\ref{sec:conclusions} summarizes the results of the paper and provides directions for future work.

\section{Methods}
\label{sec:methods}
\subsection{Model system and QED Hamiltonian}
\begin{figure}[t]
    \centering
    \includegraphics[width=\linewidth]{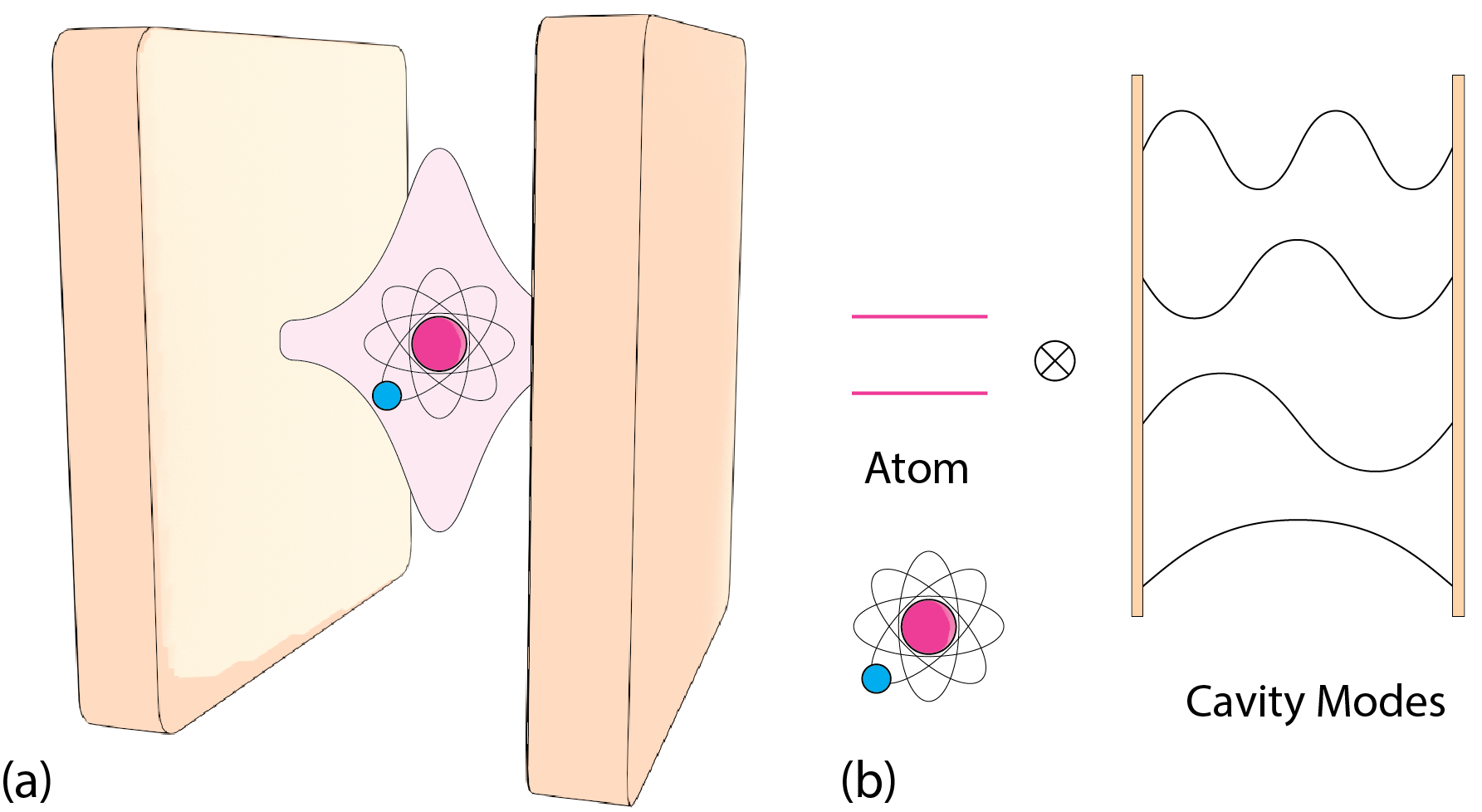}
    \caption{Schematics of a two-level fermionic system placed in the center of an optical cavity, coupled to a bath of cavity modes. a) Representation of a two-level atom placed in the center of an optical cavity. The direction of confinement is the $z$ direction. b) Representation of the coupling of the atom with the cavity modes, visualized as standing waves. Note that only the odd modes (i.e. the ones that are non-zero in the center) couple to the matter.}
    \label{fig:cartoon}
\end{figure}

To study the behavior of a generic two-level system in an optical cavity, as the one in Fig.~\ref{fig:cartoon}, we first model the two uncoupled systems (electrons and photons) and subsequently describe their interaction. All throughout this work we use atomic units.

The most efficient way to describe a two level system is using a spin representation~\cite{Miessen2021}. However, since we want to keep our theory general and extendable to multi-level systems such as in atoms and molecules, we use Fermionic operators in the form of:
\begin{equation}
    \label{eq:uncoupl_matter}
    \hat{H}_m = \sum_i \varepsilon_i \hat{c}_i^{\dagger} \hat{c}_i
\end{equation}
where $\hat{c}^{\dagger}, \hat{c}$ are electronic creation (annihilation) operators, and $\varepsilon_i$ is the energy of the i-th state. We use $\{\varepsilon_1, \varepsilon_2\} = \{-0.6738, -0.2798\}$ Ha, which corresponds to a 1D Hydrogen atom described with the soft Coulomb potential~\cite{VacuumFluctuations}.
The matter system is placed at the center of the cavity.

We describe the uncoupled light system with a set of effective harmonic oscillators:
\begin{equation}
    \label{eq:uncoupl_ph}
    \hat{H}_{ph} = \sum_{\alpha,\lambda} \Omega_{\alpha} \left(\hat{a}_{\alpha, \lambda}^\dagger \hat{a}_{\alpha, \lambda} + \frac{1}{2}\right)
\end{equation}
where $\Omega_{\alpha}$ represents the energy of the photonic mode $\alpha$ and $\lambda$ is the polarization.
We apply the long-wavelength approximation (LWA)~\cite{qed_hamiltonian} for the component of the momentum of the modes in the $xy$ plane (in-plane), meaning that $\bm{q}_{\alpha,\parallel} = \bm{0}$. Conversely, we distinguish the momenta of the different modes in the direction of confinement, hence $q_{\alpha,z} = \frac{\pi \alpha}{L}$ where $L$ is the length of the cavity. Consequently, $\Omega_{\alpha} = c q_{\alpha,z}$, where $c$ is the speed of light~\cite{Flick2017}.
We only consider one polarization ($\lambda = s$), hence we drop the index $\lambda$.

We couple light and matter by minimal coupling $\bm{\Hat{p}} \rightarrow \bm{\Hat{p} + \Hat{A}}$.
Hence, the interaction Hamiltonian reads~\cite{Flick2017, Buek1999}:
\begin{equation}
    \label{eq:h_int}
    \hat{H}_{\text{int}} = -\sum_{ij, \alpha} d_{ij} \omega_{ij} \lambda_{\alpha} \hat{c}_i^{\dagger} \hat{c}_j \hat{q}_{\alpha}
\end{equation}
where $d_{ij}$ is the dipole matrix element, $\omega_{ij}$ is the energy of the $ij$ transition, $\hat{q}_{\alpha} = \sqrt{\frac{1}{2\Omega_{\alpha}}} \left( \hat{a}_{\alpha}^{\dagger} + \hat{a}_{\alpha} \right)$ and $\lambda_{\alpha} = \sqrt{\frac{2}{L}} \sin \left(q_{\alpha, z} z\right)$ is the mode function.
Note that since the matter system is placed at the center of the cavity ($z = \frac{L}{2}$), only the odd cavity modes will have a non-zero coupling.
Despite the theory is written in general terms, in our implementation we enforce the rotating wave approximation (RWA), which simplifies the Hamiltonian by reducing the number terms, implying fewer Pauli strings $h_j$ (c.f. Eq.~\ref{eq:lie_trotter}) and fewer gates in the quantum circuit.

The full QED Hamiltonian is:
\begin{equation}
    \label{eq:h_qed}
    \hat{H}_{\text{QED}} = \hat{H}_{\text{m}} + \hat{H}_{\text{ph}} + \hat{H}_{\text{int}}
\end{equation}
Note that we absorbed the diamagnetic term into the uncoupled photon Hamiltonian by performing a Bogoliubov transformation~\cite{CavityGraphene, Rokaj2022}.

\subsection{Quantum dynamics}
To study the quantum dynamics of the two-level system in an optical cavity, we prepare the initial state such that the matter is initially in the excited state, while all cavity modes are in the vacuum.
Since this is not an eigenstate of the time-independent QED Hamiltonian in Eq.~\ref{eq:h_qed}, the system will evolve according the time-dependent Schr\"{o}dinger equation, hence: $\ket{\psi(t)} = e^{-i\hat{H}t}\ket{\psi(0)}$, where $\ket{\psi(0)}$ is the initial state of the system.
One of the most common techniques for simulating time evolution in quantum computing is approximating the unitary time-evolution operator with a product formula.
At first order, this corresponds to the Lie-Trotter formula. Given a Hamiltonian of the form $\hat{H} = \sum_{j=1}^{N_h} h_j$, that is to say that the Hamiltonian can be expressed as a sum of $N_h$ Pauli strings $h_j$, the Lie-Trotter decomposition approximates the evolution operator as
\begin{equation}
    \label{eq:lie_trotter}
    \exp(-i\hat{H}t) \approx \left( \prod_{j=1}^{N_h} e^{-i h_j \frac{t}{d}} \right)^d,
\end{equation}
where $t$ is the simulation time and $d$ is the number of time-steps.
The associated error scales as $\mathcal{O}\left(\frac{N_h t^2}{d}\right)$, as the QED Hamiltonian can be split into even and odd parts~\cite{Chiesa2019}.
The main limitation of this method lies in its circuit depth (the maximum number of layers of gates along the longest path), which tends to be large because the error scales quadratically with the total simulation time.

\subsection{Bosonic operators in \textit{Qiskit Nature}}
In order to represent both Eq.~\ref{eq:uncoupl_ph} and Eq.~\ref{eq:h_int} on a quantum computer, one needs to be able to represent bosonic operators.
To do this, we developed the class \texttt{BosonicOp} in \textit{Qiskit Nature}, representing a generic second-quantized bosonic operator.
The documentation on how to use such class is released with \textit{Qiskit Nature} (c.f. Appendix~\ref{app:qiskit_sdk} for the version details). 
In order to use such operator, one should translate it (map) to a Pauli operator. Hence, we implemented two mappers, the \texttt{BosonicLinearMapper} and \texttt{BosonicLogarithmicMapper}.
Both mappers represent a boson as a set a of Fock states $\{\ket{0}, \ket{1}, ...\}$, up to a maximum occupation $n_{\alpha}^{\text{max}}$.
The two bosonic mappers differ in how the truncated Fock space is encoded on qubits and this leads to qualitatively different qubit operators, despite the similar ladder structure of the creation operator.
In the \texttt{BosonicLinearMapper}, based on Ref.~\cite{Miessen2021}, a single mode $\alpha$ is represented by a register of $N_{\alpha,q} = n_{\alpha}^{\text{max}} + 1$ qubits in a unary code, where the occupation $n_{\alpha}$ is stored as the position of a flag along the qubit register.
For example, to represent $\{\ket{0}, \ket{1}, \ket{2}\}$ one would need three qubits, and would have $\ket{0} \rightarrow \ket{001}, \ket{1} \rightarrow \ket{010}, \ket{2} \rightarrow \ket{100}$.
Therefore, the creation operator is represented by:
\begin{equation}
    \label{eq:linear_map_create_op}
    \hat{a}_{\alpha}^\dagger = \sum_{n_{\alpha} = 0}^{n_{\alpha}^{\text{max}} - 1} \sqrt{n_{\alpha} + 1} \hat{\sigma}_{n_\alpha}^+ \hat{\sigma}_{n_\alpha + 1}^-
\end{equation}
and it shifts the flag from level $n_{\alpha}$ to $n_{\alpha + 1}$ with the proper $\sqrt{n_{\alpha} + 1}$ amplitude. $\hat{\sigma}_{n_\alpha}^\pm$ follow the usual definition of combination of Pauli matrices. The qubit count scales as $\mathcal{O}\left(N_{\alpha} \left(n_{\alpha}^{max} + 1\right)\right)$. This construction yields operators that are strictly local on the mode register (only nearest neighbors along the rail interact), so bosonic hopping terms compile to shallow circuits at the price of a linear qubit overhead per mode.

By contrast, the \texttt{BosonicLogarithmicMapper}, based on Ref.~\cite{Huang2025, Somma2003, Peng2025}, stores the occupation $n$ in a binary representation over $N_{\alpha,q} = \lceil log_2(n_{\alpha}^{\text{max}}+1) \rceil$ qubits. 
For example, to represent $\{\ket{0}, \ket{1}, \ket{2}\}$ one would need two qubits, and would have $\ket{0} \rightarrow \ket{00}, \ket{1} \rightarrow \ket{01}, \ket{2} \rightarrow \ket{10}$.
In this encoding the creation (annihilation) operator has to increase (decrease) a binary-encoded number, and it is written as:
\begin{equation}
    \label{eq:log_map_create_op}
    \hat{a}_\alpha^\dagger = \sum_{n=0}^{2^{N_{\alpha,q}} - 2} \sqrt{n_{\alpha}+1}\ \ket{n+1}_\alpha\bra{n}_\alpha
\end{equation}
The operator $\ket{n+1}_\alpha\bra{n}_\alpha$ translates, for every qubit, to a combination of Pauli matrices (c.f. Appendix~\ref{app:log_mapper}). The consequence is a reduction of the qubit count, which scales as $\mathcal{O}\left(N_\alpha \log_2\left(n_{\alpha}^{\text{max}}\right)\right)$, but at the cost of more Pauli strings and increased non-locality within the mode register, which generally leads to deeper circuits for bosonic hopping terms.

A detailed description of both mappers can be found in the Appendix~\ref{app:bosonic_mappers}.

\subsection{Mixed operators in \textit{Qiskit Nature}}
Light–matter Hamiltonians contain products of fermionic and bosonic operators (e.g. Eq.~\ref{eq:h_int}). 
To represent the corresponding joint Hilbert space in the qubit space, we developed the \texttt{MixedOp} (mixed operator) class in \textit{Qiskit Nature}, together with its mapper \texttt{MixedMapper}. A \texttt{MixedOp} is a composite second-quantized operator which aggregates heterogeneous subsystems such as fermions (\texttt{FermionicOp}) and bosons (\texttt{BosonicOp}) into a single object representing their tensor product.
In quantum-computing terms, this corresponds to forming a joint quantum register by concatenating the single registers of the single subsystems in a well-defined order.
The \texttt{MixedMapper} then delegates the qubit encoding of each operator to the specified mapper for that type for subsystem.
For instance, in this work we map the mixed operator $\hat{c}_i^\dagger \hat{c}_j \hat{a}_{\alpha}^\dagger$ by applying the Bravyi–Kitaev mapper to $\hat{c}_i^\dagger \hat{c}_j$ and the \texttt{BosonicLogarithmicMapper} to $\hat{a}_{\alpha}^\dagger$.
The \texttt{MixedMapper} automatically computes the required register lengths, and assembles the mixed qubit operator by placing each mapped subsystem at the appropriate place in the joint quantum register.
The result is a \texttt{SparsePauliOp} whose Pauli strings have length $N_j = N_{\text{reg},f} + N_{\text{reg},b}$ where $N_{\text{reg},f}$ ($N_{\text{reg},b}$) is the number of qubits in the original fermionic (bosonic) register.
This construction preserves the chosen encoding within each sector, and guarantees a consistent qubit layout across Hamiltonian terms and observables.
For instance, considering a two-level fermionic system mapped with the Bravyi-Kitaev mapper coupled to one mode represented with $\{\ket{0}, \ket{1}\}$ and mapped with the bosonic logarithmic mapper, one would have $N_{\text{reg},f} = 2$ and $N_{\text{reg},b} = 1$. Hence, the total register would have 3 qubits organized as $\ket{B_1 F_1 F_2}$, where $F_i$ is a qubit encoding the fermionic system, and $B_i$ is a qubit encoding the bosonic one.

\subsection{Noise model and hardware}
All numerical experiments are performed using \textit{Qiskit} and \textit{Qiskit Nature}.
Noise is emulated with \textit{Qiskit-Aer}, using real calibration data of the 156-qubit superconducting quantum computer \texttt{ibm\_ pittsburgh} (c.f. Appendix~\ref{app:hardware_layout} for the connectivity scheme).
The noise model includes depolarization, thermal relaxation and readout assignment errors and it represents a \textit{worst-case scenario}, as running on the quantum hardware gives access to state-of-the-art error mitigation techniques (such as dynamical decoupling and readout mitigation).
Note that due to memory and time limitations of the classical HPC cluster running the simulations, we are not able to simulate more than 24 qubits.

\section{Results}
\label{sec:results}
Here we present the results for the dynamics of the QED Hamiltonian in Eq.~\ref{eq:h_qed} for a two-level fermionic system.
We represent such matter system using the \texttt{FermionicOp} class, and we map it using the Bravyi-Kitaev mapper. As a result, we use two qubits.
We truncate the occupation of the cavity modes to $n^{\text{max}} = 1$, meaning that the Fock space for each mode is $\{\ket{0}, \ket{1}\}$. Using the bosonic logarithmic mapper, each mode is mapped to a single qubit.
The interaction terms are represented with the \texttt{MixedOp} and mapped using the \texttt{MixedMapper} (which wraps the Bravyi-Kitaev and the bosonic logarithmic mapper).
Our initial state $\ket{\psi(0)}$ is made of the matter system being in its excited state, and all cavity modes in the vacuum. Then, we let the system evolve freely.
Refer to Appendix~\ref{app:methods_values} for the numerical values.
At every time-step, we measure the occupation of the matter excited state $\hat{n}_e = \hat{c}_e^\dagger \hat{c}_e$. To perform the measurement, we use the \texttt{Estimator} primitive as implemented in \textit{QiskitAer}, which computes $\bra{\psi(t)}\hat{n}_e\ket{\psi(t)}$. In terms of Pauli operators, this corresponds to $\bra{\psi(t)}\frac{II+ZZ}{2} \otimes \mathbb{I}_{ph}\ket{\psi(t)}$, where $I, Z$ are Pauli operators and $\mathbb{I}_{ph}$ represents the identity operator applied to all qubits representing a photonic mode.

\begin{figure}[t]
    \centering
    \includegraphics[width=\linewidth]{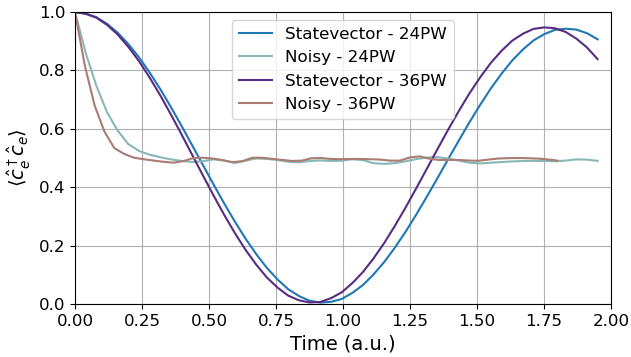}
    \caption{Quantum dynamics of a two-level system placed in the center of an optical cavity. The two-level system was initially in the excited state, while all photonic modes started in the vacuum state. For $N_{\text{ph}} = 24$ modes, the total number of qubits is 14 (2 for the matter, 12 for the photon modes). For $N_{\text{ph}} = 36$ modes, the total number of qubits is 20 (2 for the matter, 18 for the photon modes). For both cases, we report the exact statevector simulation (where we can observe a full Rabi oscillation) as well as the noisy curve (which reaches the saturation around $t \approx 0.25$ a.u.)}
    \label{fig:planewave_dynamics}
\end{figure}

\begin{figure*}[t]
    \centering
    \begin{minipage}{0.32\textwidth}
        \centering
        \fontsize{8}{8}\selectfont a) Standing-waves approach, ideal hardware
        \includegraphics[width=\linewidth]{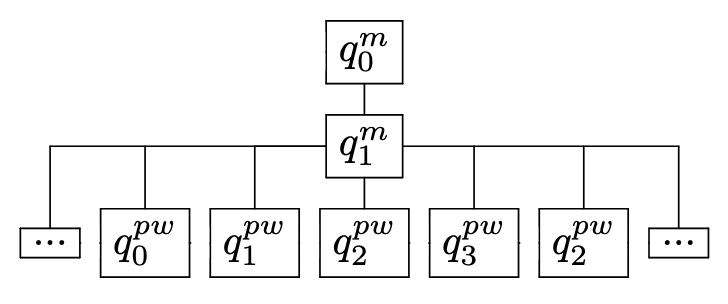}
    \end{minipage}
    \begin{minipage}{0.32\textwidth}
        \centering
        \fontsize{8}{8}\selectfont b) Localized basis approach, ideal hardware, strict $\sigma$ approximation
        \includegraphics[width=\linewidth]{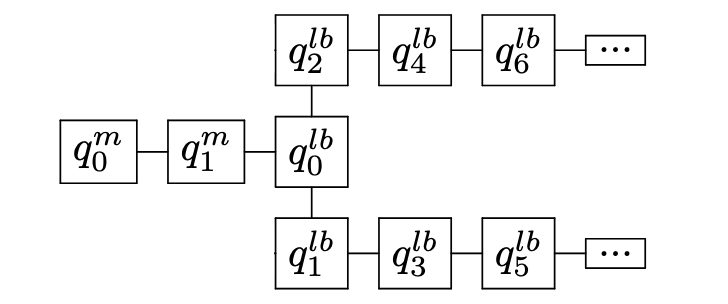}
    \end{minipage}
    \begin{minipage}{0.32\textwidth}
        \fontsize{8}{8}\selectfont c) Localized basis approach, ideal hardware, relaxed $\sigma$ approximation
        \centering
        \includegraphics[width=\linewidth]{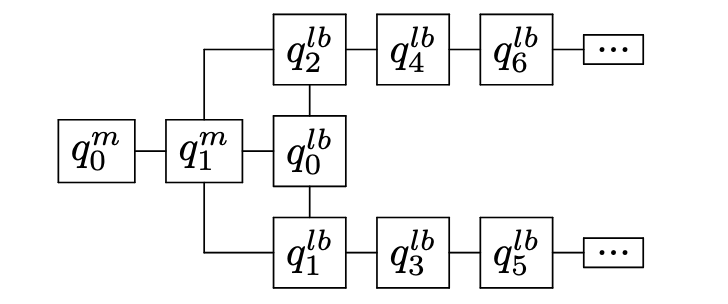}
    \end{minipage}
    \begin{minipage}{0.32\textwidth}
        \vspace{3mm}
        \centering
        \fontsize{8}{8}\selectfont d) Standing-waves approach, \texttt{ibm\_pittsburgh}
        \includegraphics[width=\linewidth]{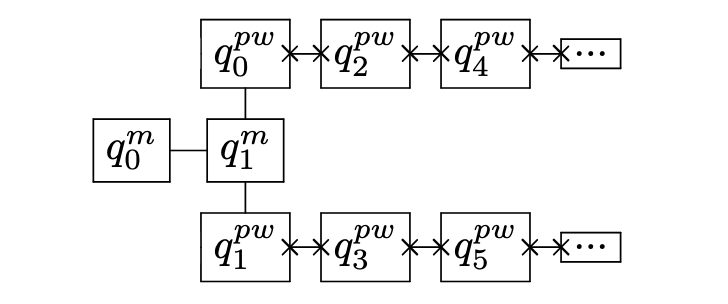}
    \end{minipage}
    \begin{minipage}{0.32\textwidth}
        \vspace{3mm}
        \centering
        \fontsize{8}{8}\selectfont e) Localized basis approach, \texttt{ibm\_pittsburgh}, strict $\sigma$ approximation
        \includegraphics[width=\linewidth]{figures/circ_layouts/lb_lin_chain.png}
    \end{minipage}
    \begin{minipage}{0.32\textwidth}
        \vspace{3mm}
        \centering
        \fontsize{8}{8}\selectfont f) Localized basis approach, \texttt{ibm\_pittsburgh}, relaxed $\sigma$ approximation
        \includegraphics[width=\linewidth]{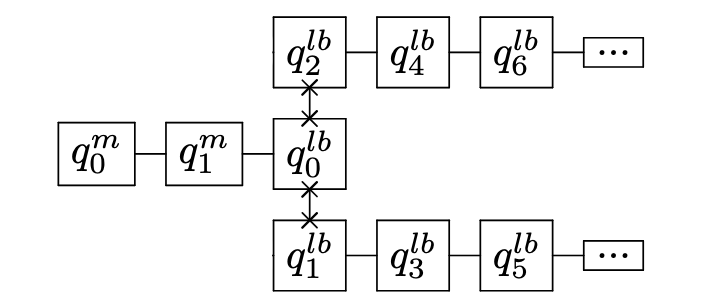}
    \end{minipage}
    \caption{Schematics of the qubit connectivity for the standing-waves approach and the localized basis approach. Panels a), b) and c) represent the connectivity for an ideal hardware, while panels d), e) and f) show the connectivity after the circuit is transpiled to \texttt{ibm\_pittsburgh}. a) Required connectivity for the standing-waves approach on an ideal hardware. All photonic qubits $q^{pw}$ are connected to the \textit{central} matter qubit $q_1^m$. b) Required connectivity for the localized basis approach on an ideal hardware, assuming that the tensor $\tau$ in Eq.~\ref{eq:h_qed_loc} is tridiagonal and $\sigma \neq 0$ only between $q_1^m$ and the central localized function $q_0^{lb}$. c) Required connectivity for the localized basis approach on an ideal hardware, assuming that the tensor $\tau$ in Eq.~\ref{eq:h_qed_loc} is tridiagonal and $\sigma \neq 0$ between $q_1^m$ and the three central localized function $q_0^{lb}, q_1^{lb}, q_2^{lb}$. d) Connectivity of the standing-waves approach mapped onto \texttt{ibm\_pittsburgh}. All of the qubits representing a mode are divided into two branches, and SWAP operations (represents by the \textit{x} in the connectors) are introduced to allow them to interact with $q_1^m$. e) Connectivity of the localized basis approach on the \texttt{ibm\_pittsburgh}. Since we enforced $\tau$ and $\sigma$ to match the hardware layout, no SWAP operations are required, which makes this panel identical to panel b). f) Connectivity of the localized basis approach on the \texttt{ibm\_pittsburgh}, when $\sigma \neq 0$ for the three central localized functions. Since the required connectivity for the qubit $q_1^m$ is 4, SWAP operations are required (in particular, 4 SWAPs per time-step between $q_0^{lb}, q_1^{lb}, q_2^{lb}$).}
    \label{fig:circ_layouts}
\end{figure*}

\subsection{Standing-waves photonic basis}
\label{subsec:planewaves}
We simulate the quantum dynamics by directly mapping the Hamiltonian in Eq.~\ref{eq:h_qed}, without additional approximations, for a two-level system initially in the excited state and all photonic modes in the vacuum state. As we describe the electromagnetic field as a superposition of standing-waves, we refer to this section as the Standing-waves approach.
Figure~\ref{fig:planewave_dynamics} shows the dynamics of the two-level system coupled to either $N_{\text{ph}} = 24$ or $N_{\text{ph}} = 36$ cavity modes.
Note that since the two level system is placed in the center of the cavity, only the odd modes couple to it.
Hence, the number of qubits in the photonic register is $N_q = 12$ or $N_q = 18$, respectively.
In both cases, we report both the noiseless (statevector) simulation and the noisy one.
While for the former we observe a full Rabi oscillation, the noisy curve quickly falls to the value of $\langle\hat{n}_e\rangle = 0.5$ and stays flat, meaning that the noise completely saturates the signal.
Note that the plateau is at $0.5$ instead of $0$ due to the term $\frac{II}{2} \otimes \mathbb{I}_{ph}$ in the observable.

The reason for such an aggressive noise is twofold.
Firstly, looking at the interaction Hamiltonian we notice that the electronic transition is coupled to all the cavity modes (through the terms $\sum_{\alpha} \hat{c}_i^\dagger \hat{c}_j (\hat{a}_{\alpha}^\dagger + \hat{a}_{\alpha})$).
This implies that one of the two qubits representing the matter system needs to interact with all the qubits of the photonic register, as depicted in Fig.~\ref{fig:circ_layouts}(a).
In the case of $N_{\text{ph}} = 24$ cavity modes, this means that one qubit has to communicate with 13 others (with 19 others for $N_{\text{ph}} = 36$).
Since the maximum connectivity on the hardware \texttt{ibm\_pittsburgh} is 3 (and in general for superconducting quantum computers it is never larger than 4), in order to allow this interaction the transpiler introduces SWAP operations, as shown in Fig.~\ref{fig:circ_layouts}(d). In particular, when $q_1^m$ has to interact with $q_2^{pw}$, first the transpiler places a SWAP between $q_0^{pw}$ and $q_2^{pw}$, then the interaction takes place and finally $q_0^{pw}$ and $q_2^{pw}$ are swapped back again.
Note that each cavity mode is independent, hence the qubits in the photonic register do not interact with each other.
The second reason is due to the high load on the central qubit. In fact, all of the CNOTs per time-step (except for the ones involved in the SWAP operations) involve the central qubit $q_1^m$, which in the case of $N_{\text{ph}} = 24$ cavity modes is 248 CNOTs (405 CNOTs for $N_{\text{ph}} = 36$).

\begin{figure*}[t]
    \centering
    \begin{minipage}{0.49\textwidth}
        \centering
        \fontsize{8}{8}\selectfont a) 24 modes, 13 localized functions, $\sigma_7 \neq 0$
        \includegraphics[width=\linewidth]{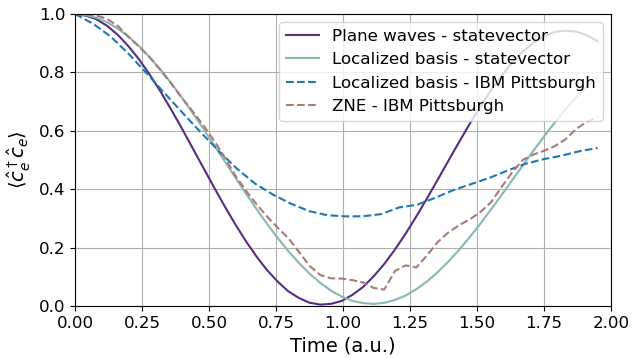}
    \end{minipage}
    \begin{minipage}{0.49\textwidth}
        \centering
        \fontsize{8}{8}\selectfont b) 36 modes, 19 localized functions, $\sigma_{10} \neq 0$
    \includegraphics[width=\linewidth]{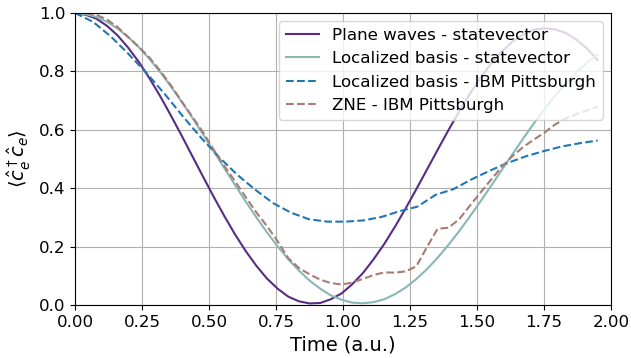}
    \end{minipage}
    \begin{minipage}{0.49\textwidth}
        \centering
        \fontsize{8}{8}\selectfont c) 24 modes, 13 localized functions, $\sigma_{6}, \sigma_{7}, \sigma_{8} \neq 0$
        \includegraphics[width=\linewidth]{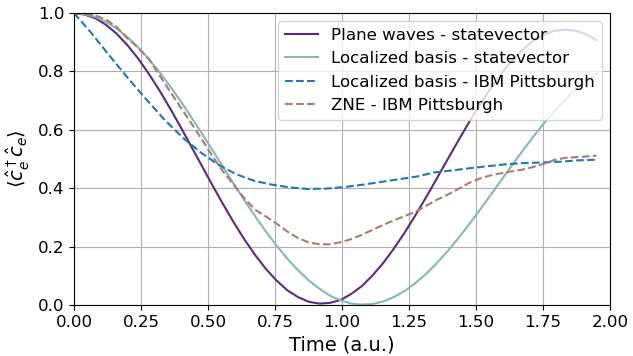}
    \end{minipage}
    \begin{minipage}{0.49\textwidth}
        \centering
        \fontsize{8}{8}\selectfont d) 36 modes, 19 localized functions, $\sigma_{9}, \sigma_{10}, \sigma_{11} \neq 0$
    \includegraphics[width=\linewidth]{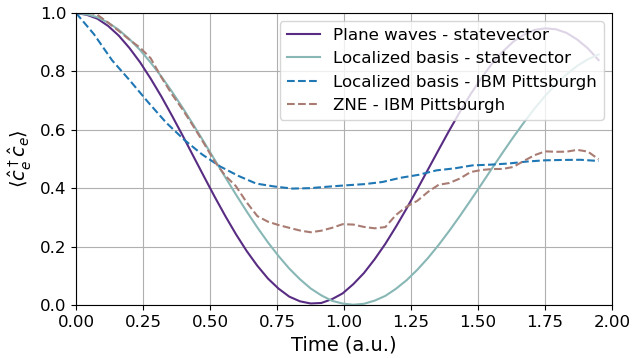}
    \end{minipage}
    \caption{Quantum dynamics of a two-level system placed in the center of an optical cavity when the modes are described with the localized basis approach (c.f. Section~\ref{sec:localizedbasis}), assuming that the tensor $\tau$ in Eq.~\ref{eq:h_qed_loc} is tridiagonal and $\sigma \neq 0$ only for a few central localized functions (1 in panels a and b, 3 in panels c and d). The two-level system was initially in the excited state, while all cavity modes started in the vacuum state. Due to the constraints on $\tau$ and $\sigma$, the statevector simulation for the localized basis does not reproduce the standing-waves reference, but still represents a good approximation. Note that relaxing the constraint on $\sigma$ leads to a better approximation for the statevector simulation, as it can be seen by comparing panel a with c, or b with d. The zero-noise extrapolation (ZNE) dynamics is obtained using a linear fit. a) $N_{\text{ph}} = 24$ cavity modes, approximated with $N_{\text{loc}} = 13$ localized functions, using $N_q = 15$ qubits. $\sigma \neq 0$ only for the central localized function ($\sigma_7$). b) $N_{\text{ph}} = 36$ cavity modes, approximated with $N_{\text{loc}} = 19$ localized functions, using $N_q = 21$ qubits. $\sigma \neq 0$ only for the central localized function ($\sigma_{10}$). c) $N_{\text{ph}} = 24$ cavity modes, approximated with $N_{\text{loc}} = 13$ localized functions, using $N_q = 15$ qubits. $\sigma \neq 0$ for the three central localized functions ($\sigma_6, \sigma_7, \sigma_8$). d) $N_{\text{ph}} = 36$ cavity modes, approximated with $N_{\text{loc}} = 19$ localized functions, using $N_q = 21$ qubits. $\sigma \neq 0$ for three central localized functions ($\sigma_{9}, \sigma_{10}, \sigma_{11}$).}
    \label{fig:loc_basis_dynamics}
\end{figure*}

\subsection{Localized photonic basis}
\label{sec:localizedbasis}
To address the connectivity bottlenecks, we perform a unitary transformation of the photonic modes with the aim of making the Hamiltonian representable as a 1D linear chain of qubits (so that no SWAP operation is needed).
To achieve that, we consider a set of localized and orthogonal basis functions. By \textit{localized} we mean that the function is peaked around a specific point inside the cavity and then quickly goes to zero (c.f. Appendix~\ref{app:basis_func}).
We project the photonic operators onto this new basis:
\begin{equation}
    \label{eq:loc_basis_op}
 \hat a_\alpha        = \sum_{l=0}^\infty P_{l \alpha}\,\hat{t}_l, \qquad
 \hat a_\alpha^\dagger= \sum_{l=0}^\infty P^*_{l \alpha}\, \hat{t}_l^\dagger,
\end{equation}
where $P_{l \alpha}=\int L_l(z)\,e^{iq_{\alpha, z}z} \mathrm{d}z\;$ is the projection of the $\alpha$-th mode onto the $L_l(z)$-th basis function.
The operators $\hat{t}^\dagger, \hat{t}$ create (annihilate) an excitation in the specified localized basis function and, importantly, we treat them as bosons.
In this work, we use a set of triangular basis functions (c.f. Appendix~\ref{app:basis_func} for the exact definition), but others may be employed as well (i.e. rectangular).
Expanding the Hamiltonian in Eq.~\ref{eq:h_qed} with the operators in Eq.~\ref{eq:loc_basis_op}, we obtain (c.f. Appendix~\ref{app:loc_basis_derivation} for the complete derivation):
\begin{equation}
    \label{eq:h_qed_loc}
    \begin{aligned}
        & \hat{H}_{\text{QED}}^{\text{loc}} = \hat{H}_m + \sum_{\alpha}^{N_{ph}}\frac{\Omega_\alpha}{2} + \\
        & \sum_{ll'}^{N_{loc}} \tau_{ll'} \hat{t}_l^\dagger \hat{t}_{l'} - \sum_{ij} d_{ij} \omega_{ij} \sum_{l}^{N_{loc}} \left(\sigma_l^* \hat{t}_l^\dagger + \sigma_l \hat{t}_l\right) \hat{c}_i^\dagger \hat{c}_j
    \end{aligned}
\end{equation}
where $\hat{H}_m$ is the matter Hamiltonian defined in Eq.~\ref{eq:uncoupl_matter}, $i,j$ are indexes that span over the matter states and $\omega_{ij}$ is the energy of the matter transition. $N_{\text{loc}}$ is the number of localized functions used in the expansion.
The tensor $\tau_{ll'} = \sum_{\alpha} \Omega_{\alpha} P_{l\alpha}^* P_{l'\alpha}$ represents the hopping between the localized functions $l$ and $l'$, while $\sigma_l = \sum_{\alpha} \lambda_\alpha \sqrt{\frac{1}{2\Omega_{\alpha}}} P_{l,\alpha}$ represents (together with $d_{ij} \omega_{ij}$) the probability amplitude of having a transition between the state $i$ and $j$ through the destruction of a localized excitation at $l$.
As the operators $\hat{t}^\dagger, \hat{t}$ have bosonic nature, we represent them using the \texttt{BosonicOp} class and, limiting their occupation to the state $\ket{1}$, they require only one qubit to be encoded on the quantum computer.
The localized Hamiltonian in Eq.~\ref{eq:h_qed_loc} requires, in general, a much worse connectivity than the Hamiltonian in Eq.~\ref{eq:h_qed} due to the fact that the interaction term behaves in the same way (coupling each matter transition to all the localized functions), and in addition the hopping tensor $\tau$ requires an all-to-all connectivity between the qubits representing the localized functions.
In order for the Hamiltonian to be representable as a linear chain, the tensor $\tau$ should be tridiagonal (meaning that an excitation can only hop to the neighboring basis function) and $\sigma$ should be non-zero only for one localized function.
If these two conditions are met, the required qubit connectivity becomes the one shown in Fig.~\ref{fig:circ_layouts}(b, e), where a qubit has to interact at most with three neighboring qubits (which is supported by the honeycomb layout of \texttt{ibm\_pittsburgh}).
In Appendix~\ref{app:limitations} we discuss the limitations of such approximations.

Fig.~\ref{fig:loc_basis_dynamics} shows the quantum dynamics of the two-level system coupled to the cavity modes described using the localized basis approach. We focus on panels (a, b), where $\tau$ and $\sigma$ are truncated to meet the aforementioned conditions.
First, we note that the ideal (statevector) simulation of the localized basis approach, while still qualitatively capturing the physics of the Rabi oscillations, does not quantitatively reproduce the statevector simulations of the standing-waves approach (which is our reference).
This is expected as we use a finite set of localized functions and we truncate both $\tau$ and $\sigma$ to meet the connectivity requirements.
Such a result represents a good trade-off between algorithmic fidelity (i.e. how well we can reproduce the reference result) and noise resilience.
In fact, thanks to the improved connectivity, the noisy curve reaches the noise saturation value of $\langle \hat{n}_e \rangle \approx 0.5$ for $t > 2$ a.u., compared to the standing-waves approach where it was reached at $t \approx 0.25$ a.u.
Furthermore, the noisy curve in Fig.~\ref{fig:loc_basis_dynamics}(b) shows that even when expanding the localized basis set (hence, when using more qubits) the noise remains almost constant. In fact, Fig.~\ref{fig:loc_basis_dynamics}(b) uses six qubits more than Fig.~\ref{fig:loc_basis_dynamics}(a), but the noisy curve follows that same dynamics.
This resilience is a major improvement as it allows to apply the zero-noise extrapolation (ZNE) technique~\cite{Li2017, Endo2018} to mitigate the quantum error. In the ZNE one deliberately amplifies the error from the quantum gates and then extrapolates the zero-noise curve. We only amplify the two-qubits gates as they are responsible for the majority of the noise. We do that by substituting a single two-qubit gate with three two-qubits gates, which in a noiseless environment does not modify the simulation as the application of two consecutive two-qubits gates is equal to the identity (c.f. Appendix~\ref{app:error_mitigation} for more details).
By applying such a technique, we can quantitatively reconstruct the noiseless dynamics up to half of the Rabi oscillation, and only qualitatively in the second part of the oscillation.
This is a striking result because it shows that algorithms based on product formulas (such as Trotter), which are usually considered a bad choice for near-term quantum dynamical application due to their circuit depth, can be successfully used to simulate cavity-QED systems.

\subsection{Beyond the linear chain approximation}
\label{sec:beyond_lin_chain}
In the previous section we enforced a strictly 1D connectivity by truncating the hopping tensor $\tau$ to a tridiagonal form and allowing the coupling vector $\sigma$ to be non-zero only for the central localized function. While this choice maximizes hardware compatibility, it reduces algorithmic fidelity with respect to the standing-waves reference. We now relax the linear-chain constraint.

In the following, we keep $\tau$ tridiagonal but allow $\sigma$ to be non-zero for the three central localized functions. Then, the connectivity scheme becomes the one shown in Fig.~\ref{fig:circ_layouts}(c).
In this scenario, the matter qubit $q_1^m$ requires connections with 4 other qubits, which exceeds the maximum connectivity on \texttt{ibm\_pittsburgh}. Hence, the transpiler introduces SWAP operations, as shown in Fig.~\ref{fig:circ_layouts}(f). In particular, the circuit will have two SWAPs per time-steps between $q_0^{lb}$ and $q_1^{lb}$, and other two between $q_0^{lb}$ and $q_2^{lb}$.

Fig.~\ref{fig:loc_basis_dynamics}(c,d) shows the quantum dynamics for $N_{\text{ph}} = 24$ and $N_{\text{ph}} = 36$ modes, approximated with $N_{\text{loc}} = 13$ and $N_{\text{loc}} = 19$ localized functions, respectively. 
Comparing the statevector simulations for the localized basis approach between Fig.~\ref{fig:loc_basis_dynamics}(a) and Fig.~\ref{fig:loc_basis_dynamics}(c) (or between Fig.~\ref{fig:loc_basis_dynamics}(b) and Fig.~\ref{fig:loc_basis_dynamics}(d)), we notice that relaxing the constraint on $\sigma$ leads to a better agreement with the reference standing-waves curve. In general, the amount of the improvement will depend on the ratio between the coupling of the central function $\sigma_{\text{central}}$ and the coupling of the neighboring localized functions $\sigma_{\text{neighbor}}$. In all panels of Fig.~\ref{fig:loc_basis_dynamics} we have $\frac{\sigma_{\text{neighbor}}}{\sigma_{\text{central}}} < 20\%$, hence the improvement is small.
On the contrary, comparing the noisy curve between Fig.~\ref{fig:loc_basis_dynamics}(a) and Fig.~\ref{fig:loc_basis_dynamics}(c) (or between Fig.~\ref{fig:loc_basis_dynamics}(b) and Fig.~\ref{fig:loc_basis_dynamics}(d)) we notice that the overhead of SWAP operation has a significant impact, as the noisy curve reaches the noise saturation value of $\langle \hat{n}_e \rangle \approx 0.5$ at $t \approx 1.5$ a.u., compared to the Fig.~\ref{fig:loc_basis_dynamics}(a, b) where saturation happened after $t = 2$ a.u. 
Despite this expected worsening, the noisy simulations remain substantially more robust than in the standing-waves approach (where the noise saturation was reached at $t \approx 0.25$ a.u.), and, more importantly, still allows the application of the ZNE technique. 
We apply it following the procedure described in Sec.~\ref{sec:localizedbasis}, and we recover the short-time dynamics with high accuracy up to $t \approx 0.7$ a.u. (about 10 time-steps). Beyond this time, the extrapolated curve starts to deviate more noticeably as errors compound across the deeper circuit.
Yet, it still tracks the qualitative behavior of the reference localized basis statevector simulation.
Moreover, given the fast development of quantum hardware, one can expect it will soon be possible to retrieve the quantitative behavior as well (c.f. Appendix~\ref{app:future_hardware}).

\section{Conclusions and outlook}
\label{sec:conclusions}
In this work we have introduced a novel approach for simulating cavity-QED Hamiltonians on near-term quantum processors.
First, we presented the implementation of the \texttt{BosonicOp} and \texttt{MixedOp} classes, together with their respective mappers, in \textit{Qiskit Nature}. These allow to represent a QED Hamiltonian, which comprises of both pure fermionic and bosonic operators and mixed operators, on a quantum computer.
Subsequently, we used them to study the quantum dynamics of a two-level system in an optical cavity, using two approaches (both based on a Trotterized Hamiltonian).

We find that directly encoding the QED Hamiltonian (which we call standing-waves approach) fails on noisy hardware due to its star-like connectivity. In fact, due to the nature of the interaction Hamiltonian, one of the qubits representing the matter system has to interact with all the qubits that represent a cavity mode.
To overcome this, we propose a novel approach where we perform a unitary transformation on the QED Hamiltonian and represent the modes with a localized photonic basis, such that the matter only interacts with one basis function and each site only interacts with its nearest neighbor. This allows to represent the Hamiltonian as a 1D chain of qubits, hence matching the hardware topology and avoiding SWAP operations. This enables error mitigation via zero-noise extrapolation to recover the quantum dynamics up to half the Rabi oscillation.
Furthermore, we showed that the noise scaling on this localized approach is robust enough that we can still recover a significant portion of the quantum dynamics when we relax the 1D constraints on the qubit connectivity, even in the presence of SWAP operations.
Despite not being able to exactly reproduce the standing-waves reference, these findings show that product-formula time evolution, when combined with hardware-aware mappings, are effective for QED simulations even on near-term devices.
Considering the fast development of the quantum hardware, one can assume that in the next few years, with the advent of fault-tolerant devices, it will be possible to retrieve the full Rabi oscillation even when the Hamiltonian is not representable as a linear chain (c.f. Appendix~\ref{app:future_hardware}).

Future work should point to different directions. Firstly, we focused on a two level system, which only allows a description of simple models. One should consider a multi-level molecule.
On the algorithmic side, one should try to improve the definition of the localized functions, such that the conditions on the hopping tensor $\tau$ and on the interaction coefficient $\sigma$ are met with a softer truncation. 
Alternatively, one could further relax the requirements on those quantities, especially on $\sigma$, to improve the algorithmic fidelity.
This will be especially interesting once fault-tolerant superconducting quantum computers will be released in the next few years or on other architectures (i.e. ion-trap, neutral atoms ...) which naturally have an all-to-all qubit connectivity.

\section*{Data availability}
The data that support the findings of this study is available from the corresponding authors upon reasonable request.

\section*{Code availability}
The code that support the findings of this study is available from the corresponding authors upon reasonable request.

\section*{Acknowledgements}
The authors sincerely thank Anthony Gandon for his precious work in the implementation of the \texttt{MixedOp} and \texttt{MixedMapper} classes in \textit{Qiskit Nature}.
We also thank Max Rossmannek for his contribution in the design of the classes implemented in \textit{Qiskit Nature}.
This research was supported by NCCR MARVEL, a National Center of Competence in Research, funded by the Swiss National Science Foundation (grant number 205602) and by RESQUE funded by the Swiss National Science Foundation (grant number 225229).
We acknowledge support from the Villum foundation grant No. 72146, the Cluster of Excellence "CUI: Advanced Imaging of Matter" - EXC 2056 - project ID 390715994, European Research Council (ERC-2024-SyG-101167294; UnMySt) and Grupos Consolidados (IT1453-22), and the Max Planck-New York City Center for Non-Equilibrium Quantum Phenomena. We acknowledge support from the European Union Marie Sklodowska-Curie Doctoral Networks TIMES grant No. 101118915 and SPARKLE grant No. 101169225. The Flatiron Institute is a division of the Simons Foundation.

\section*{Author contribution}
F.T. derived the theory, developed all the code (except for the \texttt{MixedOp} and \texttt{MixedMapper} classes, to which he contributed with technical design and code review), performed all simulations, analyzed the data, interpreted the results and wrote the paper.
S.L. contributed to the derivation of the theory, to the interpretation of the results and partly to the writing of the paper.
M.L. and H.A. contributed to the derivation of the theory and to the interpretation of the results.
I.T. contributed to the derivation of the theory, to the analysis of the data and to the interpretation of the results.
A.R. contributed to the interpretation of results.
A.R. and I.T. conceived the project.
All authors contributed to the revision of the manuscript.

\section*{Competing Interests}
All authors declare no financial or non-financial competing interests.

\bibliography{references}
\bibliographystyle{quantum}

\onecolumn
\appendix

\section{Outlook for future hardware}
\label{app:future_hardware}
In this appendix we repeat the simulations for the quantum dynamics of a two-level system when the cavity modes are approximated with the localized basis approach (c.f. Fig.~\ref{fig:loc_basis_dynamics} of the main text), comparing different generations of quantum hardware.
Since in Sec.~\ref{sec:localizedbasis} and Sec.~\ref{sec:beyond_lin_chain} we showed that this approach is resilient to noise when increasing the number of localized function, here we reproduce only Fig.~\ref{fig:loc_basis_dynamics}(a,c), corresponding to 24 modes approximated with 13 localized functions.
We compare the performance of the following hardware: \texttt{ibm\_brisbane} (127 superconducting qubits, with QPU \textit{Eagle r3}), \texttt{ibm\_pittsburgh} (156 superconducting qubits, with QPU \textit{Heron r3}) and a custom superconducting hardware, representing a future development of IBM's QPUs.
This latter was modeled such that it has the same layout of \texttt{ibm\_pittsburgh} (c.f. Fig.~\ref{fig:ibm_pittsburgh}), but with a significantly better noise.
Particularly, we accessed the real-time calibration data of \texttt{ibm\_pittsburgh} and we decreased the average one- and two-qubit gate errors, $e^{\text{pitts}}_1$ and $e^{\text{pitts}}_2$, respectively, as well as readout errors of the device, $e^{\text{pitts}}_{\text{read}}$, while simultaneously increasing average relaxation time $T_1$ and dephasing time $T_2$, by a factor $\eta$~\cite{Miessen2021}:
\begin{equation}
    e^{\text{custom}}_1 = \frac{e^{\text{pitts}}_1}{\eta}, \quad e^{\text{custom}}_2 = \frac{e^{\text{pitts}}_2}{\eta}, \quad e^{\text{custom}}_{\text{read}} = \frac{e^{\text{pitts}}_{\text{read}}}{\eta}, \quad T^{\text{custom}}_1 = \eta T^{\text{pitts}}_1, \quad T^{\text{custom}}_2 = \eta T^{\text{pitts}}_2
\end{equation}
We used $\eta=10$. The resulting data (for all hardware) are summarized in Table~\ref{tab:qpu_specs}.

\begin{figure*}[t]
    \centering
    \begin{minipage}{0.49\textwidth}
        \fontsize{8}{8}\selectfont a) 24 modes, 13 localized functions, $\sigma_7 \neq 0$
        \centering
        \includegraphics[width=\linewidth]{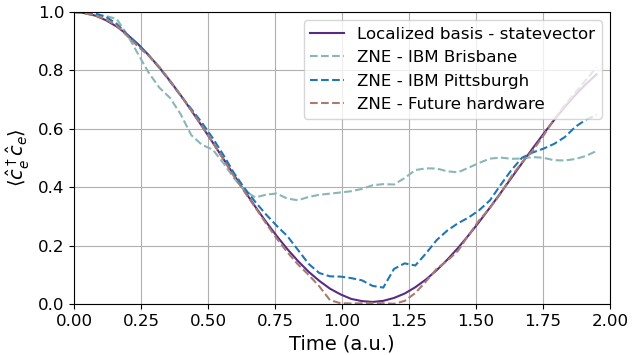}
    \end{minipage}
    \begin{minipage}{0.49\textwidth}
        \fontsize{8}{8}\selectfont b) 24 modes, 13 localized functions, $\sigma_6, \sigma_7, \sigma_8 \neq 0$
        \centering
        \includegraphics[width=\linewidth]{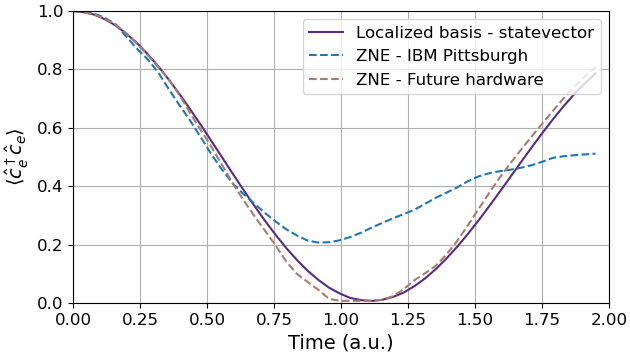}
    \end{minipage}
    \caption{Quantum dynamics of a two-level system placed in the center of an optical cavity when the modes are described with the localized basis approach, assuming that the tensor $\tau$ in Eq.~\ref{eq:h_qed_loc} is tridiagonal and $\sigma \neq 0$ only for a few central localized functions (1 in panel a , 3 in panel b). The two-level system was initially in the excited state, while all cavity modes started in the vacuum state. The ZNE dynamics is obtained using a linear fit. $N_{\text{ph}} = 24$ cavity modes, approximated with $N_{\text{loc}} = 13$ localized functions, using $N_q = 15$ qubits. $\sigma \neq 0$ only for the central localized function ($\sigma_7$). b) $N_{\text{ph}} = 24$ cavity modes, approximated with $N_{\text{loc}} = 13$ localized functions, using $N_q = 15$ qubits. $\sigma \neq 0$ for the three central localized functions ($\sigma_6, \sigma_7, \sigma_8$).}
    \label{fig:hardware_comparison}
\end{figure*}

The results are presented in Fig.~\ref{fig:hardware_comparison}. 
Panel (a) compares the performances of \texttt{ibm\_brisbane}, \texttt{ibm\_pittsburgh} and the custom hardware when $\sigma$ is non-zero only for one localized function, and the Hamiltonian is represented as a linear chain of qubits (c.f. Fig.~\ref{fig:circ_layouts}(b,e)).
\texttt{ibm\_brisbane} is able to reproduce the noiseless curve only up to $t \approx 0.7$ a.u., while the other QPUs make it possible to reconstruct the full Rabi oscillation (i.e. they do not reach the noise saturation within the plotted interval).
In the case of the custom QPU, the noiseless curve and its noisy equivalent are almost indistinguishable (except when $\langle \hat{n}_e\rangle \rightarrow 0$).
Panel (b) compares the performances when $\sigma$ is non-zero only for three central localized function, hence the "beyond the linear chain approximation" case discussed in Sec.~\ref{sec:beyond_lin_chain} (c.f. Fig.~\ref{fig:circ_layouts}(c,f) for the required connectivity).
We did not include \texttt{ibm\_brisbane} because of its poor performance in panel (a) (the noise in this case is much bigger).
While \texttt{ibm\_pittsburgh} is able to reproduce the noiseless curve only up to $t \approx 0.75$ a.u., the custom hardware retrieves the full Rabi oscillation (despite the presence of deviations).

This result confirms what we stated in the main text, showing that product-formula time evolution, when combined with hardware-aware mappings, are effective for QED simulations, especially when considering the hardware that will be available in the next few years.

\begin{table}[t]
    \centering
    \caption{QPU Specifications and Performance Metrics as of \textit{September 10$^{\text{th}}$, 2025}}
    \resizebox{\textwidth}{!}{%
    \begin{tabular}{|l|c|c|c|c|c|c|c|c|c|c|c|c|c|c|c|}
        \hline
        \textbf{Name} & \textbf{Qubits} & \textbf{QPU Type} & \textbf{Gates} & \textbf{2Q Layer} & \textbf{CLOPS} & \textbf{Read Err} & \textbf{2Q Err} & \textbf{SX Err} & \textbf{T1 ($\mu$s)} & \textbf{T2 ($\mu$s)} \\
        \hline
        \textit{Brisbane} & 127 & Eagle r3 & ecr, id, rz, sx, x & $1.72\times10^{-2}$ & 180K & $2.05\times10^{-2}$ & $6.76\times10^{-3}$ & $2.49\times10^{-4}$ & 222.44 & 133.23 \\
        \hline
        \textit{Pittsburgh} & 156 & Heron r3 & cz, id, rx, rz, rzz, sx, x & $4.14\times10^{-3}$ & 250K & $4.33\times10^{-3}$ & $1.52\times10^{-3}$ & $1.80\times10^{-4}$ & 296.33 & 357.45 \\
        \hline
        \textit{Custom} & 156 & N/A & cz, id, rx, rz, rzz, sx, x & $4.14\times10^{-4}$ & 250K & $4.33\times10^{-4}$ & $1.52\times10^{-4}$ & $1.80\times10^{-5}$ & 2960.33 & 3570.45 \\
        \hline
    \end{tabular}
    }
    \label{tab:qpu_specs}
\end{table}

\section{Numerical methods}
\subsection{Qiskit SDK}
\label{app:qiskit_sdk}
To perform the simulations we used the following versions of Qiskit SDK Python packages:
\texttt{qiskit}: v1.4.4, \texttt{qiskit-aer}: v0.17.1, \texttt{qiskit-ibm-runtime}: v0.41.1, \texttt{qiskit-nature}: cloned from \textit{main} branch of the repository (commit SHA: 4cc927ce539219505defac61cd70ded081507361).

\subsection{Hardware layout}
\label{app:hardware_layout}
\begin{figure}
    \centering
    \includegraphics[width=0.5\linewidth]{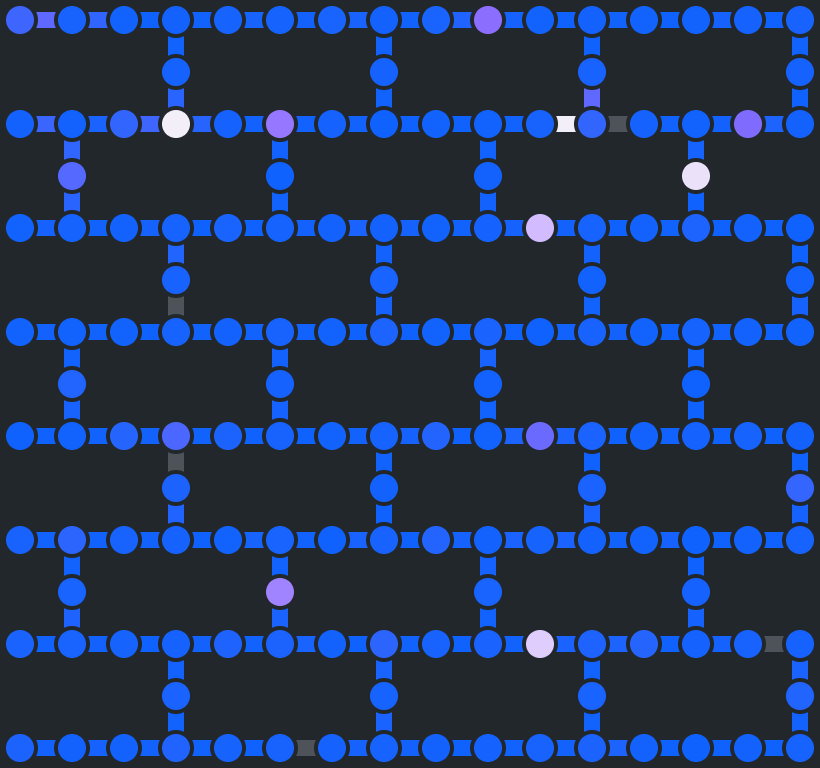}
    \caption{QPU layout of \textit{IBM Pittsburgh}. The color of each qubit gives information about its readout error, while the color of each connection represents the error of a \textit{cz} operation between the connected qubits. In both cases, the color blue represents a small error probability, the color white a large error.}
    \label{fig:ibm_pittsburgh}
\end{figure}

We emulated the hardware \texttt{ibm\_pittsburgh} on a classical HPC cluster. At the time of writing, this is IBM's flagship's hardware.
Its QPU (of type \textit{Heron r3}) has 156 superconducting qubits, organized in a 2D honeycomb lattice (c.f. Fig.~\ref{fig:ibm_pittsburgh}).
The majority of the qubits is connected to 2 other qubits, while the highest connectivity is 3.
Refer to Table~\ref{tab:qpu_specs} for more details.

\subsection{Noise simulation \& mitigation}
\label{app:error_mitigation}
The noise model was simulated by accessing the real-time calibration data of the QPU through the Python package \texttt{qiskit-ibm-runtime}, and built using the \texttt{NoiseModel} class implemented in the Python package \texttt{qiskit-aer}.
In order to obtain the noisy simulations (in Fig.~\ref{fig:planewave_dynamics} and Fig.~\ref{fig:loc_basis_dynamics}), we ran each circuit 10 times, and then averaged.
The required precision of the Estimator job was set to $10^{-4}$.

In order to mitigate the error, we used the zero-noise extrapolation (ZNE) technique. 
To perform it, we first simulated the circuit as it was transpiled (for a total of 10 simulations). Averaging these 10 simulations, we obtained the unamplified noisy curve (which is reported in Fig.~\ref{fig:loc_basis_dynamics}, as the dashed blue curve).
Subsequently, we executed a circuit where $10\%$ of the two-qubits gates was amplified (again, we performed a total of 10 simulations and then averaged). This way, we obtained the $10\%$ amplified noise curve.
The gate amplification was performed by substituting the desired two-qubits gate with 3 two-qubits gates. Since the application of two consecutive two-qubits gates is equivalent to an identity gate, on a noiseless hardware this operation does not change the outcome. However, on a noisy QPU it amplifies the error probability.
We then repeated same amplification procedure, increasing the percentage to $20\%, 30\%, 40\%, 50\%, 60\%$ of the total two-qubits gates.
To extract the ZNE curve, we then interpolated the averaged curves with a linear interpolation, obtaining the dashed brown line in Fig.~\ref{fig:loc_basis_dynamics}.

\subsection{Time evolution \& other parameters}
\label{app:methods_values}
We simulate the time evolution by Trotterizing the QED Hamiltonian. To contain the number of two-qubits gates we use the Lie-Trotter formula, which is a first order approximation.
The chosen time-step is $\Delta t = 0.075$ a.u. and the total duration is $t_f = 2.0$ a.u., for a total of 26 time steps.

We set the dipole matrix element in Eq.~\ref{eq:h_int} to $d_{eg} = 60$ a.u., which is an arbitrary value that allows us to observe a full Rabi oscillation in total simulated time. The energy of the electronic transition of the two-level system is $\omega_{eg} = 0.394$ Ha.
When we simulate $N_{\text{ph}} = 24$ modes we set the cavity length to $L = 13000$ a.u., while when we simulate $N_{\text{ph}} = 36$ modes we set it to $L = 19500$ a.u.. This ensures that half of the effective modes are below $\omega_{eg}$, and the other half are above that value (ensuring a symmetrical cavity).

\section{Localized basis}
\label{sec:loc_basis}
In this appendix we provide more information on the localized basis approach described in Section~\ref{sec:localizedbasis} of the main text.

\subsection{Derivation}
\label{app:loc_basis_derivation}
First, we show how to obtain the localized QED Hamiltonian (Eq.~\ref{eq:h_qed_loc} of the main text).
Note that this derivation will only assume the new basis to be orthogonal, but not localized. Hence, it is general to a larger class of basis functions.
Let us define the expansion of the photonic creation and annihilation operators onto the new basis as:
\[
 \hat a_\alpha        = \sum_{l=0}^\infty P_{l \alpha}\,\hat{t}_l, \qquad
 \hat a_\alpha^\dagger= \sum_{l=0}^\infty P^*_{l \alpha}\, \hat{t}_l^\dagger,
\]
where $P_{l \alpha}=\int L_l(\bm{r})\,e^{i\bm{q}_{\alpha}\cdot \bm{r}} \mathrm{d}\bm{r}\;$ is the projection of the $\alpha$-th mode onto the $l$-th basis function.
If the cavity is planar, such as in this work, then the integral $P_{l \alpha}$ is only along the confinement direction (in our case $z$).
Note that such transformation is unitary in the infinite limit (i.e. $\alpha \rightarrow \infty, l \rightarrow \infty$).
Substituting the projections of the photonic creation and annihilation operators into the QED Hamiltonian in Eq.~\ref{eq:h_qed} leads to:
\begin{equation}
    \hat{H}_{\text{QED}} = \hat{H}_m + \sum_{\alpha} \Omega_{\alpha} \left(\frac{1}{2} + \sum_{ll'} P_{l\alpha}^* P_{l'\alpha} \hat{t}_l^\dagger \hat{t}_{l'}\right) - \sum_{ij}d_{ij} \omega_{ij} \hat{c}_i^\dagger \hat{c}_j \sum_{\alpha} \lambda_{\alpha} \sqrt{\frac{1}{2\Omega_{\alpha}}} \sum_l \left( P_{l\alpha}^* \hat{t}_l^\dagger + P_{l\alpha} \hat{t}_l \right) 
\end{equation}
where $\hat{H}_m = \sum_i \varepsilon_i \hat{c}_i^{\dagger} \hat{c}_i$ is the matter Hamiltonian defined in Eq.~\ref{eq:uncoupl_matter}, $i,j$ are indexes that span over the matter states, $\omega_{ij}$ is the energy of the matter transition $ij$ and $d_{ij}$ is its dipole matrix element.
Finally, reordering the previous expression and introducing the quantities:
\begin{equation}
    \tau_{ll'} = \sum_\alpha \Omega_\alpha P_{l\alpha}^* P_{l'\alpha}, \quad \sigma_l = \sum_\alpha \lambda_{\alpha} \sqrt{\frac{1}{2\Omega_{\alpha}}} P_{l\alpha}
\end{equation}
leads to:
\begin{equation}
    \hat{H}_{\text{QED}}^{\text{loc}}= \hat{H}_m + \sum_{\alpha} \frac{\Omega_{\alpha}}{2} + \sum_{ll'} \tau_{ll'} \hat{t}_l^\dagger \hat{t}_{l'} - \sum_{ij, l} d_{ij} \omega_{ij} \hat{c}_i^\dagger \hat{c}_j \left( \sigma_{l}^* \hat{t}_l^\dagger + \sigma_{l} \hat{t}_l \right) 
\end{equation}

\subsection{Basis functions of choice}
\label{app:basis_func}
We choose a set of localized basis functions $L_l(z)$ defined as:
\begin{equation}
    L_l(z) =
    \begin{cases}
        1 - m_l|z - z_{0,l}|, & z \in \left[-\frac{1}{m_l} + z_{0,l}, \frac{1}{m,l} + z_{0,l}\right] \\
        0, & \text{otherwise}
    \end{cases}
\end{equation}
which represents a triangle centered at $z_{0,l}$ with steepness $m_l$.
We define such functions such that they cover all points in the planar cavity (which goes from $z=0$ to $z=L$). To achieve that, we define $m_l = \frac{2N_{\text{loc}}}{L}$, where $N_{\text{loc}}$ is the number of localized basis functions and $z_{0,l} = \frac{1}{m_l} \left(1 + l\frac{L}{N_{\text{loc}}} \right)$. This definition ensures the orthogonality of the basis functions.
We then normalize each function such that $\langle L_l(z) | L_{l'}(z) \rangle = \delta_{ll'}$, hence obtaining an orthonormal set.
Such definition is ideal because if $N_{\text{loc}}$ is odd, then one function will peak at the center of the cavity ($z=\frac{L}{2}$), which is where the two-level system is located.

\begin{figure*}[t]
    \centering
    \begin{minipage}{0.49\textwidth}
        \fontsize{8}{8}\selectfont a)
        \centering
        \includegraphics[width=\linewidth]{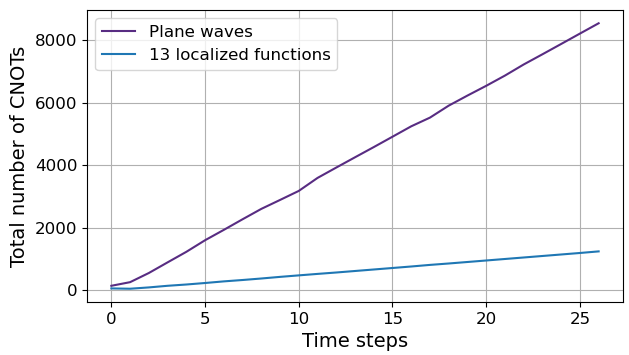}
    \end{minipage}
    \begin{minipage}{0.49\textwidth}
        \fontsize{8}{8}\selectfont b)
        \centering
        \includegraphics[width=\linewidth]{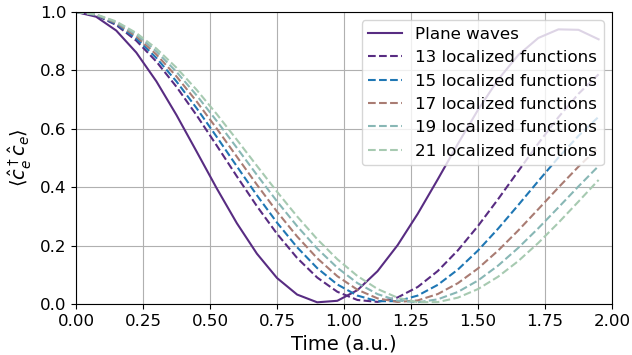}
    \end{minipage}
    \caption{The two panels were obtained for the two-level system in the main text coupled to 24 cavity modes. a) Evolution of the total number of CNOTs for the standing-waves approach and for the localized basis approach when $\tau$ is tridiagonal and $\sigma$ is non-zero only for the central function. b) Noiseless quantum dynamics of the two-level system.}
    \label{fig:loc_approx}
\end{figure*}

\subsection{Limitations}
\label{app:limitations}
The limitations of the localized standing-waves approach are mainly to the hardware constraints. In the following we only consider a two level system, hence we drop the indexes $i,j$ in the localized QED Hamiltonian.
Each basis function (i.e. each operator $\hat{t}$) requires one qubit to be represented on the quantum computer, and the connectivity is determined by the shape of the hopping tensor $\tau$ and by $\sigma$.
As we discuss in the main text, if $\tau$ is tridiagonal (i.e. an excitation can only hop to nearest neighbor localized function) and $\sigma$ is non-zero only for the central localized function, then the connectivity scheme is the one pictured in Fig.~\ref{fig:circ_layouts}(c). Then, $\hat{H}_{\text{QED}}^{\text{loc}}$ can be mapped to the hardware without any need for SWAP operations and the number of total CNOTs improved significantly compared to the standing-waves approach, as shown in Fig.~\ref{fig:loc_approx}.
Relaxing such conditions would result in the need of SWAP operations. For instance, if one allows $\sigma$ to be non zero for the central three localized function , then the qubit $q_{1}^{m}$ would need to interact with $q_{0}^{lb}, q_{1}^{lb}, q_{2}^{lb}$. To achieve this, the transpiler would need to first make $q_{1}^{m}$ and $q_{0}^{lb}$ interact. Subsequently, swap $q_{1}^{lb}$ and $q_{0}^{lb}$, make $q_{1}^{m}$ and $q_{0}^{lb}$ and re-swap $q_{1}^{lb}$ and $q_{0}^{lb}$ (and then follow the same procedure for $q_{0}^{lb}$).
Aside from introducing many noisy two-qubits gates, they would all be localized on a few physical qubits, strongly increasing the error probability.
Hence, the two conditions on $\sigma$ and $\tau$ shall be met.

However, given the localized functions described in Appendix~\ref{app:basis_func}, the two conditions are never met. Thus, one has to manually truncate $\tau$ and $\sigma$.
First, one should always choose an odd number of basis functions, as with an even number of them there would be two $\sigma$ with the same values. This would make the manual truncation ineffective and force to introduce SWAP operations.
Secondly, due to the need to enforce the conditions on $\tau$ and $\sigma$, one has an optimal number of localized function for which the approximation is best.
Fig.~\ref{fig:loc_approx}(b) shows that using more localized functions results in a worse approximation of the standing-waves reference, even in a noiseless simulation.

\section{Bosonic operator mappers}
\label{app:bosonic_mappers}
This appendix details how the bosonic mappers implemented in \textit{Qiskit Nature} work.

\subsection{The linear mapper}
The \texttt{BosonicLinearMapper} is based on Section II.C of Ref.~\cite{Miessen2021}. As the name suggests, the creation and annihilation operators for the mode $k$ are linearly mapped to the qubit space, and they are defined as follows:
\begin{align}
    \label{eq:linear_map_create}
    & \hat{b}_k^\dagger = \sum_{n_k = 0}^{n_k^{max} - 1} \sqrt{n_k + 1} \hat{\sigma}_{n_k}^+    \hat{\sigma}_{n_k + 1}^-\\
    \label{eq:linear_map_annihilate}
    & \hat{b}_k = \sum_{n_k = 0}^{n_k^{max} - 1} \sqrt{n_k + 1} \hat{\sigma}_{n_k}^- \hat{\sigma}_{n_k + 1}^+
\end{align}
where $n_k^{max}$ is the maximum occupation of the mode and $k$ is the index of the mode. The operators $\hat{\sigma}_{n_k}^-$ and $\hat{\sigma}_{n_k}^+$ are a combination of Pauli matrices:
\begin{align}
    \label{eq:sigma_+}
    & \hat{\sigma}_{n_k}^+ = \frac{1}{2} \left(\hat{\sigma}_{n_k}^x + i\hat{\sigma}_{n_k}^y\right) \equiv S_{n_k}^+ = \frac{1}{2} \left(X_{n_k} + iY_{n_k}\right) \\
    \label{eq:sigma_-}
    & \hat{\sigma}_{n_k}^- = \frac{1}{2} \left(\hat{\sigma}_{n_k}^x - i\hat{\sigma}_{n_k}^y\right) \equiv S_{n_k}^- = \frac{1}{2} \left(X_{n_k} - iY_{n_k}\right)
\end{align}
In order to represent a mode we need $n_k^{max} + 1$ qubits (in other words, for a single mode the length of the qubit register is $n_k^{max} + 1$). 
In general, the memory complexity of this mapper (i.e. how many qubits it requires) is given by $\mathcal{O}\left(N_K n_k^{max}\right)$, where $N_K$ is the number of bosonic modes. Hence, the scaling is linear with both the number of modes and the maximum allowed occupation per mode.
One should note that since this mapper truncates the maximum occupation of a bosonic mode to represent it in the qubit register, the commutation relations of the mapped operator differ from the standard ones.
Please refer to Section 4, equation 22 of Ref.~\cite{Somma2003} for more details.
This essentially implies that we have a further fundamental relation of the bosonic operator~\cite{Veis2016}:
\begin{equation}
    \label{eq:creation_zero}
    \hat{b}^\dagger |n_k^{max}\rangle = 0
\end{equation}
which prevents exceeding the maximum representable occupation.

Let us look at a generic occupation number vector, which gives a representation of the physical qubits in the register.
Every entry can be either $0 (\uparrow)$ or $1 (\downarrow)$.
For a generic system, with $K$ modes and $n^{max}$ Fock states per mode (all modes have the same maximum occupation), this would look like:
\begin{equation}
    \label{eq:linear_occ_vector}
    \begin{aligned}
        &|m_K, m_{K-1}, ..., m_1, m_0\rangle = \\
        &|\underbrace{0_{n^{max}}, 0_{n^{max} - 1}, ..., 0_1, 0_0}_{\text{mode $K$}}, \underbrace{0_{n^{max}}, 0_{n^{max} - 1}, ..., 0_1, 0_0}_{\text{mode $K-1$}}, ..., \underbrace{0_{n^{max}}, 0_{n^{max} - 1}, ..., 0_1, 0_0}_{\text{mode $1$}}, \underbrace{0_{n^{max}}, 0_{n^{max} - 1}, ..., 0_1, 0_0}_{\text{mode $0$}} \rangle
    \end{aligned}
\end{equation}
In \textit{Qiskit}, the right-most entry represents the least significant qubit (little-endian convention).
Note that when an operator is mapped, it has to perform an action (which could be an identity operation) on every qubit in the register.

\subsection{The logarithmic mapper}
\label{app:log_mapper}
Mapping a bosonic mode has many similarities with representing an integer number on a classical computer.
In fact, contrary to the case of Fermions where a state can be either occupied or unoccupied, an infinite number of Bosons can occupy the same quantum state (which is explained by the Bose-Einstein statistics).
Therefore, representing a bosonic mode only requires the inclusion of enough Fock states so that the number of bosons occupying it becomes representable.
Since the computational basis of bits and qubits is the same, this problem is not different than representing integers on classical bits.
Thus, it is natural to try to encode the qubit register in the same way the classical register is encoded. To do this, we shall consider the representation of unsigned integers:
\[0 = 00; |0\rangle = |0, 0\rangle = |\uparrow, \uparrow\rangle\]
\[1 = 01; |1\rangle = |0, 1\rangle = | \uparrow, \downarrow\rangle\]
\[2 = 10; |2\rangle = |1, 0\rangle = |\downarrow, \uparrow\rangle\]
\[3 = 11; |3\rangle = |1, 1\rangle = |\downarrow, \downarrow\rangle\]
On the left, the representation of the unsigned integers on two classical bits is reported, while on the right the representation of the Fock states on two qubits.
In general, for $N_q$ qubits one has \cite{Peng2025}:
\begin{align*}
    & |0\rangle = |0_{N_q}, 0_{N_q - 1}, ..., 0_{1}, 0_0\rangle \\
    & |1\rangle = |0_{N_q}, 0_{N_q - 1}, ..., 0_{1}, 1_0\rangle \\
    & |2\rangle = |0_{N_q}, 0_{N_q - 1}, ..., 1_{1}, 0_0\rangle \\
    & |3\rangle = |0_{N_q}, 0_{N_q - 1}, ..., 1_{1}, 1_0\rangle \\
    & ...
\end{align*}
\begin{equation}
    \label{eq:log_state}
    |2^{N_q} - 1\rangle = |1_{N_q}, 1_{N_q - 1}, ..., 1_{1}, 1_0\rangle
\end{equation}
Again, we use the little-endian convention so the right-most entry is the least significant qubit.
Using Eq.~\ref{eq:log_state} and the well-known bosonic creation and annihilation relations, it is now possible to define what the action of the mapped operator should be \cite{Peng2025, Veis2016}:
\begin{align}
    \label{eq:log_map_create}
    & \hat{b}^\dagger = \sum_{n = 0}^{2^{N_q} - 2} \sqrt{n + 1} |n+1\rangle \langle n|\\
    \label{eq:log_map_annihilate}
    & \hat{b} = \sum_{n = 1}^{2^{N_q} - 1} \sqrt{n} |n - 1\rangle \langle n|
\end{align}
where $\ket{n}$ represents a generic Fock state. In order to define the effect of the operator in terms of the Pauli matrices, let us consider an example with $N_q = 2$ qubits.
Then, using Eq.~\ref{eq:log_state} to expand Eq.~\ref{eq:log_map_create} we get:
\begin{equation}
    \label{eq:log_2_qubit_create}
    \hat{b}^\dagger = \sum_{n = 0}^{2^{2} - 2} \sqrt{n + 1} |n+1\rangle \langle n| = |1\rangle\langle 0| + \sqrt{2}|2\rangle\langle 1| + \sqrt{3}|3\rangle\langle 2| = |0,1\rangle\langle 0,0| + \sqrt{2}|1,0\rangle\langle 0,1| + \sqrt{3}|1,1\rangle\langle 1,0|
\end{equation}
As one can see, after using the state vector to expand the Eq.~\ref{eq:log_map_create} and Eq.~\ref{eq:log_map_annihilate}, the outer operations are reduced to only four cases.
These correspond to all possible combinations of the states where the physical qubit can exist \cite{Peng2025}:
\begin{equation}
    |0\rangle \langle 0| = \frac{\mathcal{I} + \sigma^z}{2}, \quad |1\rangle \langle 1| = \frac{\mathcal{I} - \sigma^z}{2}, \quad |0\rangle \langle 1| = \sigma^+ = \frac{\sigma^x + i\sigma^y}{2}, \quad |1\rangle \langle 0| = \sigma^- = \frac{\sigma^x - i\sigma^y}{2}
\end{equation}

Let us now compare mapping the operator $\hat{b}^\dagger$ (or $\hat{b}$) for a system composed of a single mode with the logarithmic and with the linear mapper, assuming that the maximum occupation of the mode is $n_k^{max} = 3$.
For the linear mapper, one needs $n_k^{max} + 1 = 4$ qubits and 12 Pauli terms in the mapped Hamiltonian.
For the logarithmic mapper, on the other hand, one only needs $\log_2\left(n_k^{max} + 1\right) = 2$ qubits. Also, the number of terms in the mapped operator is reduced to 8. This implies that the circuit will be shallower.

Finally, let us explore what happens to the logarithmic mapper for a generic case of $N_K$ boson modes.
In Eq.~\ref{eq:linear_occ_vector} we discussed the generic state vector for $N_K$ modes and $n^{max}$ Fock states per mode for the linear mapper.
In the state $|m_K, m_{K-1}, ..., m_1, m_0\rangle = |m_K,\rangle \otimes |m_{K-1}\rangle \otimes ... \otimes |m_1\rangle \otimes |m_0\rangle$ each mode can be occupied with a different number of particles due to the tensor product between the modes. This implies each single-mode register should be independent.
Since the physical object is the same (i.e. the bosonic creation or annihilation operator), the behavior of the logarithmic mapper should be analogous, so the global register still requires $N_K$ single-mode registers.
Thus, using a logarithmic mapper instead of the linear mapper only affects the fact that fewer qubits are needed to represent the generic mode $k$.
Therefore, the complexity of the logarithmic mapper will scale linearly with respect to the number of modes, and logarithmically to the number of Fock states per mode: $\mathcal{O}\left(N_K \log_2\left(n_k^{max}\right)\right)$.

As a consequence, we can extend Eq.~\ref{eq:log_map_create} and Eq.~\ref{eq:log_map_annihilate} to the multi-mode case by adding a mode index:
\begin{align}
    \label{eq:log_map_create_multi}
    & \hat{b}_k^\dagger = \sum_{n = 0}^{2^{N_q^k} - 1} \sqrt{n + 1} |n+1\rangle_k \langle n|_k\\
    \label{eq:log_map_annihilate_multi}
    & \hat{b}_k = \sum_{n = 1}^{2^{N_q^k} - 1} \sqrt{n} |n - 1\rangle_k \langle n|_k
\end{align}
where $N_q^k$ is the number of qubits used to represent the mode $k$.
Finally, one should note that using the logarithmic mapper instead of the linear one, while always ensuring the usage of less qubits, may require more Pauli strings. For instance, this is the case for a hopping bosonic term: $\hat{b}_1^\dagger \hat{b}_2$.

\end{document}